# Towards a more realistic citation model: The key role of research team sizes

Staša Milojević [1]

[1] Center for Complex Networks and Systems Research, Luddy School of Informatics, Computing, and Engineering, Indiana University, Bloomington; smilojev@indiana.edu

**Abstract:** We propose a new citation model which builds on the existing models that explicitly or implicitly include "direct" and "indirect" (learning about a cited paper's existence from references in another paper) citation mechanisms. Our model departs from the usual, unrealistic assumption of uniform probability of direct citation, in which initial differences in citation arise purely randomly. Instead, we demonstrate that a two-mechanism model in which the probability of direct citation is proportional to the number of authors on a paper (team size) is able to reproduce the empirical citation distributions of articles published in the field of astronomy remarkably well, and at different points in time. Interpretation of our model is that the intrinsic citation capacity, and hence the initial visibility of a paper, will be enhanced when more people are intimately familiar with some work, favoring papers from larger teams. While the intrinsic citation capacity cannot depend only on the team size, our model demonstrates that it must be to some degree correlated with it, and distributed in a similar way, i.e., having a power-law tail. Consequently, our team-size model qualitatively explains the existence of a correlation between the number of citations and the number of authors on a paper.

**Keywords:** citation model; cumulative advantage; preferential attachment; team science

## 1. Introduction

Citation of scientific articles plays an important role in the contemporary ecology of science, revealing its cognitive structure (e.g., [1, 2]), and serving as a basis for research evaluation [3], despite a growing criticism [4]. Different aspects of referencing behavior, which gives rise to citations, have been extensively studied, as well as the meaning of citations and practices for their inclusion [5-8]. Publications receive citations at different rates, even when works are published in the same journal and at the same time, resulting in wide citation distributions [9]. One characteristic of the citation distribution that has been attracting much attention ever since the seminal work by Price [10] is the existence of the power-law tail. The power-law tail is a strong sign of an underlying inequality, and has led Price, in the first model of citation dynamics, to propose cumulative advantage as its generative process [11]. According to the cumulative advantage process "future accumulation depends upon current accumulation" [12] (p. 273). Cumulative advantage is also known as a preferential attachment, especially in the context of the growth of complex networks [13].

Many studies have focused on the empirical determination of a functional form that mathematically describes a citation distribution, proposing, among others, the power-law (at least in the tail) [14], log-normal [15-17] and shifted power-law functions [18]. A few studies [19, 20] have derived the functional forms analytically, starting from a citation model. However, the question of the functional form is separate from the development of the citation model, the latter focusing on uncovering the processes that lead to the empirical citation distribution, or potentially some other empirical property involving citations. Furthermore, we point out that finding mathematical (or network) processes that lead to the empirical distribution is not the same as uncovering the basis for this process, i.e., identifying the actual *citing mechanisms*, rooted in citing behavior. In this study we will focus on both aspects of this question. Citation models from the network perspective have been reviewed in references [21, 22]. Here we present a summary of key developments and their interpretations.

Based on the hypothesis presented in [10], Price [11] proposed a citation model where the accumulation process is a linear function of the existing number of citations. The linear cumulative advantage, as opposed to a power-law dependent one, has been confirmed more recently by Jeong et al. [23] and Redner [17]. In the strict cumulative advantage (CA) model papers with zero citations can never acquire any, what Price called a ground-state problem [11, 24]. To ameliorate this initiation problem, Price has departed from a strict CA model, adding a constant (one) to the citation count, while realizing that the solution is tentative and a somewhat arbitrary ("a fudge factor"). This limitation was not critical since Price's model focused on explaining the tail of the distribution.

The next important development came more than two decades later, and started a renewed interest in citation models, especially from the network perspective. Redner [25] was the first to propose that citation mechanisms driving citation accumulation of rarely cited and highly cited papers are different. He did not name the mechanisms, but said that the rarely cited papers have shorter citation life and are mostly cited by the author(s) and "close associates" (p. 132), while highly cited papers "become known through collective effects" (p. 132), without specifying what they were. Based on these considerations, Krapivsky and Redner [26] proposed a two-mechanism model called the "growing network with redirection" (GNR), in which a new paper can either cite some paper, or instead cite a reference from that paper, in a process they call "redirection". The tendency of authors to learn about papers in that way and "copy" references from other papers has been supported by Wu and Holme [27], who identified a large number of triangles in the citation network, i.e., cases where paper A cites papers B and C, but C was already cited by B, i.e., B redirects A to C. It should be pointed out that the mechanism of learning about other works through their reference lists and subsequently citing them, which than leads to cumulative advantage, is a more natural interpretation than the interpretation often given in the network literature, that papers with many citations will somehow "attract" more new citations. Similar sentiment was expressed by Redner [17].

The two mechanisms proposed by Krapivsky and Redner [26] correspond to what Peterson et al. [20] call "direct" and "indirect" citing mechanisms, the terms that we adopt as well. Peterson et al. [20] found that two mechanisms are necessary to describe the entire citation distribution and not only the power-law tail. Peterson et al.'s two-mechanism model is mathematically equivalent to Krapivsky and Redner's redirection graph network. Based on the model, Peterson et al. also provide an analytical function for the citation distribution that depends on two parameters (probability of indirect citation and the average number of references per paper).

Some citation dynamics models attempted to also account for citation age effects. For example, Eom and Fortunato [18] in their model incorporated the effect that older literature is in general cited less frequently (obsolescence). They found the "linear preferential attachment with time dependent initial attractiveness", which is essentially a two-mechanism model, to reproduce the empirical data well. The initial attractiveness of the paper is defined as "appeal to attract edges, regardless of degree" (p. 3), and it corresponds to the direct citation mechanism. Golosovsky and Solomon [28] proposed a stochastic dynamic model based on self-exciting point process, in which citation growth follows a slightly superlinear process and depends on the most recent performance of each paper. Apart from the population-level models discussed above, Wang et al. [29] proposed a three-factor model of citation accumulation of individual papers, with three factors being: preferential attachment, aging, and fitness or quality.

While previous works on citation models have led to significant progress, as measured for example by a good agreement between model and empirical distributions, almost all of these models contain one critical limitation - they assume that the probability of direct citation is *uniform*, whether they call it attractiveness, initial attractivity, or some other name. This means that in the initial period different papers accumulate citation basically randomly, and some will have more citations than the other simply by chance. Having this arbitrariness in the model is problematic, because these early citations are directly responsible for future trajectories that will be greatly amplified by the process of cumulative advantage. Price [11] did not propose an actual citation mechanism that would lead to the cumulative advantage. He did, however, point out the importance of the initial "pulse" of citations that forms the basis for subsequent accumulation, and suggested that it may be related to paper's "quality". There is no reason to assume that this initial pulse will follow a random (Poisson) distribution as it does in current models. The only model that to our knowledge considers non-uniform probability of direct citation is [18], where they instead assume that the probability is drawn from some power-law distribution, whose connection with any measurable properties is, however, missing.

In this work, we propose that one characteristic of contemporary science, the increased team effort, may also play a role in allowing co-authored papers much needed initial *visibility* that creates non-uniformity in the probability of direct citation and leads to different levels of citation advantage. Namely, a larger number of authors per paper means that more individuals will be aware of a paper's existence, leading to increased capability to ensure higher visibility [30]. That this sort of direct citing mechanism may be an important element of a citation model lies in the empirical fact that the papers written by larger teams are more likely to attract citations [31].a feature that no citation model so far explains. In this study we will use common science-of-science and scientometrics operationalization of teams as co-authors on a paper. However, we are aware that co-authorship may only be a partial indicator of true teams [32].

We base this study on the premise that considering realistic scenarios as to how people cite papers paves the way to a successful citation model. Also, many previous studies used datasets comprising of heterogeneous research areas and sometimes with unequal citation windows. Keeping these two factors constant in this paper will allow us to focus solely on citation mechanisms.

The specific goals of the present paper are to: (a) use simulations to test in detail various citation models against homogeneous empirical data at multiple times; (b) provide interpretation of Price's citation model in the context of two citation mechanisms, (c) search for an alternative to the unrealistic uniform probability of direct citation used in previous models. In our analysis we will also model the number of papers with no citations which were neglected in previous models. We recognize that the questions of the mathematical description of the citation distribution and the citation dynamics (i.e., obsolescence, bursts, etc.) are independent from the citation mechanisms and do not focus on them.

## 2. Materials and Methods

*2.1. Data selection*

The goal of this study is to investigate fundamental citation mechanisms by comparing different models (simulations) with the real data. It is therefore preferable to test the models with an empirical dataset that would minimize the influence of disciplinary differences and obsolescence, two factors identified by previous studies as having significant effect on citations and their distributions [22]. The field of astronomy represents a particularly good choice, because it is fairly active (large publication volumes) and yet the majority of research articles is published in only several journals with the same breadth of topics, shared audience, and similar citation patterns. We use Web of Science Core Collection with updates through the end of 2017 to select all items classified as "articles" from the four core journals (*Astronomy & Astrophysics*, *Astronomical Journal*, *Astrophysical Journal and Monthly Notices of the Royal Astronomical Society*) published in one calendar year (2007). These four journals have almost identical journal Impact Factors (between 5.4 and 5.8 in 2019 edition of JCR), so our analysis will not be affected by the differences in the visibility of the venues. We chose a relatively short publication window so that all articles have had a similar amount of time to accrue citations. We chose a relatively distant publication window (a decade) so that we can follow the citation dynamics long after the initial period. There are 6430 articles in this dataset. This is the only empirical dataset that has been considered in the course of this study and it was chosen prior to any modeling was attempted.

Next, we identify all instances in which an item (a document of any type published in any venue) cites any of these 6430 articles. There are 263,371 such citation instances coming from 99,691 items published between 2007 and 2017. In Figure 1 we show the distribution of citation instances. Citation reached a peak two years after the articles were published, followed by the drop of the total number of citations. The drop of citing (article obsolescence) has been relatively modest, reaching 55% of the peak 9 years after the articles were published. In the final year there was even an increase in citation.

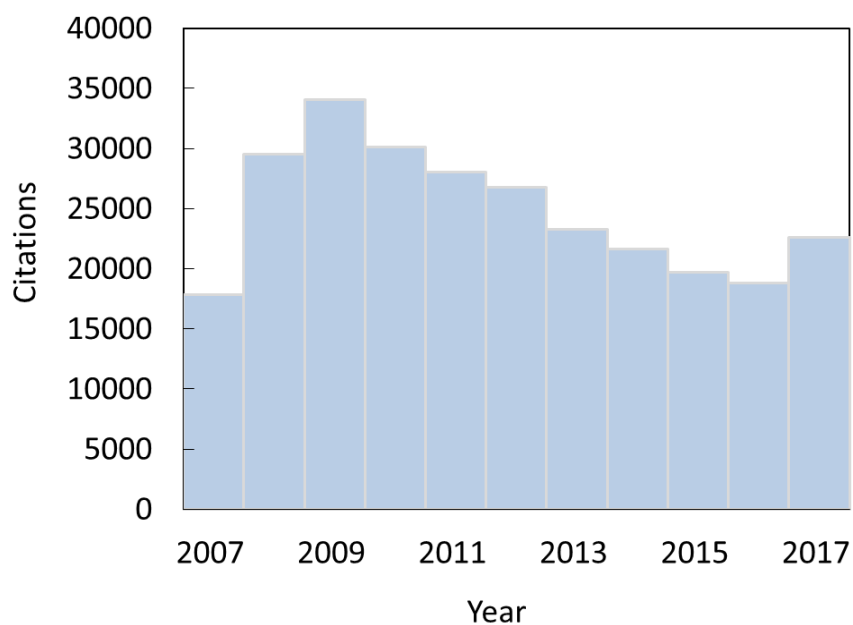

**Figure 1.** Annual citation rates of 6,430 astronomy articles published in 2007. Obsolescence is relatively modest. The number for 2007 has been multiplied by two to reflect the fact that articles had, on average, 6 months over which they could be cited.

*2.2. Citation distribution*

The key empirical property that the models have a goal of reproducing is the *citation distribution*. At the end of 2017, each article has received, on average, 41 citations. This number hides the fact that the distribution of citations is very wide. Some 39 articles received no citations at all, whereas 13 articles received more than 500, with the maximum number being 2042. Final raw citation distribution (black dots) is shown in a log-log plot in Figure 2. It reveals the presence of a "fat tail" that is typical for citation distributions, but also for many other publication metrics (distribution of total productivity, distribution of team sizes). Such tails are also called the power-law tails, even though they can deviate significantly from a pure power law that would look like a straight line on Figure 2. Performing the binning following the procedure of [33] allows us to better characterize the far tail of the distribution (red line), and therefore more stringently test model predictions. The binned distribution continues the curving trend present in the near tail.

Figure 2 adds one to the number of citations in order to be able to display the publications that have not received any citations. We see that these articles naturally follow the trend set by the articles that have more than zero citations. Given enough time, most citation distributions will acquire a peak at citations greater than zero. For this dataset this happens already after a year, as shown by binned citation distribution for citations received by the end of 2008, which peaks for papers with one citation. As the time progresses and citations accumulate (grey lines) the peak shifts to larger values. At the final time, the peak is relatively broad - similar number of articles have between 6 and 18 citations. In many datasets studied previously the peak was at lower number of citations, either because of the short citation window or because the majority of journals included in those datasets had low impact.

The benchmark for our modeling efforts will be to successfully reproduce both the final (2017) and the initial (2008) citation distributions. The two distributions primarily differ in the position and the shape of the peak region.

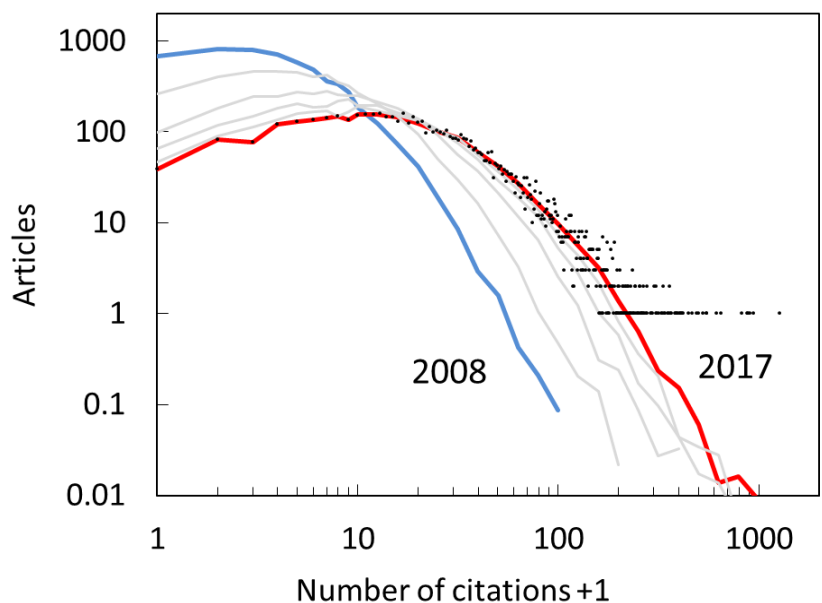

Figure 2. Evolution of the citation distribution of the empirical dataset. Distribution is shown on a log-log scale for years 2008 (blue line), years 2009, 2011, 2013 and 2015 (gray lines) and 2017 (red line and black dots). Lines represent binned averages in order to reveal the far tails of distributions. Distributions include articles with no citations by adding one to the x axis.

*2.3. Modeling methodology*

We test different citation models by performing simulations of citation "instances". In the simulations, we follow the accumulation of citations to each of 6430 articles - the number of articles in the empirical dataset. We produce 263,371 citation instances, the number that corresponds to the total citations received by 2017. Which article receives a citation in the simulations will follow the probability, which depends on the citation model and mechanism(s). Time is not featured explicitly in the simulations, but the correspondence with the actual time can be made by observing the simulations after a certain number of citation events. Specifically, to get the initial citation distribution we carry out 38,414 citation events, the number that corresponds to the total number of citations accumulated by the end of 2008.

**3. Results**

*3.1. Direct and indirect citation mechanisms*

Unlike many works that approach citation from the (general) network perspective, we wish to build a model and provide interpretation for it from the standpoint of how researchers actually cite scientific literature, i.e., by understanding citation mechanisms. We argue that the direct and indirect citation mechanisms are indeed fundamental, because they correspond to how researchers encounter the work of others, and this awareness of the existence of a particular paper is a necessary step preceding each act of referencing. Direct mechanism includes scenarios by which a researcher learns about a paper by means other than seeing it cited in some other work. Direct citation may result from diverse activities that include searches for relevant articles, learning about some work from colloquia or seminars, and, importantly, from being involved in a study, either as an author or collaborators, a colleague, or as a reviewer of the work. However, direct citation mechanisms do not in general case lead to a cumulative advantage. This requires an indirect mechanism, which includes any process by which a researcher learns about some work because it was cited in some other work. This other work is usually a publication, but it can in principle be a reference given in a presentation or some other form of scholarly communication. Both mechanisms contribute to a paper's citation accumulation throughout its citation lifecycle, but with the passage of time, indirect mechanism becomes dominant and is necessary for the outstanding citation success of certain papers.

*3.2. Uniform probability model*

In order to demonstrate the need for a two-mechanism model, and specifically the one in which the probability of direct citation is non-uniform, it is useful to consider and describe the citation distributions resulting from other, conceptually simpler models. We start with a model that contains only one mechanism: citing with uniform probability. This is a direct citing mechanism because the probability of citing does not depend on previous citations. Each paper has the same chance of being cited. We can mathematically describe this model as the probability ($P$) of citing of some paper to be:

$$P \propto 1 \qquad (1)$$

Such model would correspond to reality if all papers had the same *citing potential* and the same *visibility*, and each citing instance is entirely unaffected by any previous citation. Differences in citation counts between different papers would still be present and arise by chance, but we expect the citation distribution to be narrow.

The outcome of the uniform probability model after the first full year ("initial") and at the end of the period under investigation (~9 years, "final") is shown In Figure 3, along with the empirical citation corresponding to these times. Modeled and empirical distributions are very different in terms of the width and the location of the peak. Modeled distributions peak strongly around the mean number of citations. It can be shown that they follow Poisson distributions. For the final distribution the mean citation rate is high (~40), so the Poisson distribution is very close to a normal one. There are no papers with fewer than ~18 or more than ~65 citations in it.

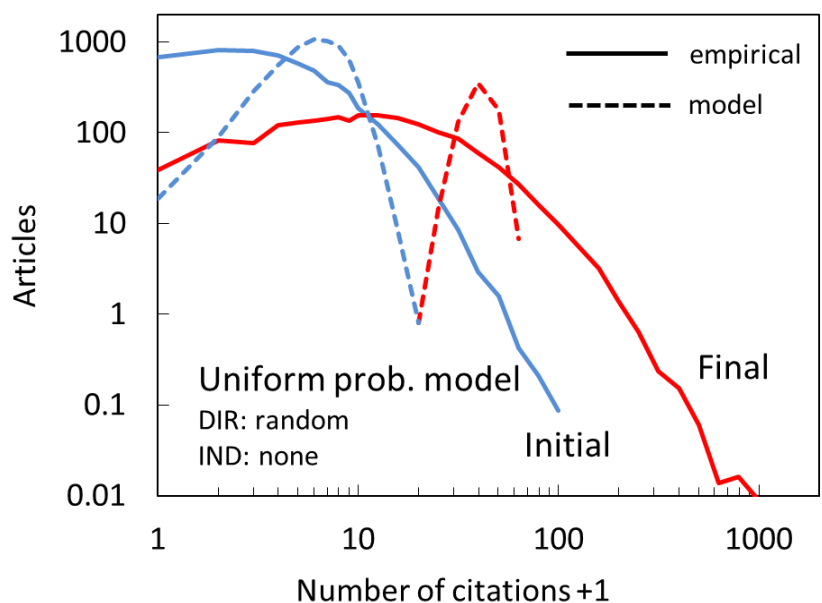

Figure 3. Citation distribution for uniform probability model (dashed lines) compared to empirical distributions (full lines) after the first full year (Initial, blue) and at the end of the full period (Final, red), corresponding to years 2008 and 2017. In this model there is only a direct citation mechanism with a uniform probability.

*3.3. Cumulative advantage model*

A citation model that lies at the heart of citation modeling is the cumulative advantage (CA) model. Following our definition, CA is an indirect mechanism. The most basic CA scenario is the one in which the probability of citing depends directly on the existing number of citations:

$$P \propto n_{cit}. \qquad (2)$$

This model makes sense from the perspective of a citation mechanism: researchers learn about other papers when they see them cited. Whereas one can generalize CA to be any monotonically increasing function of citations ($P \propto f(n_{cit})$), the simplest form (direct proportionally) is the most natural because every citation translates directly and proportionally into increased visibility and therefore increased probability of new citation.

In Figure 4 we show simulation results for a model based on this simplest form of CA. There is no direct citation in this model. The resulting distribution follows the power-law distribution throughout, and not just in the tail, as expected for a pure CA processes [26]. Apparently, a model based only on CA does not come close to the empirical distributions at any point in time.

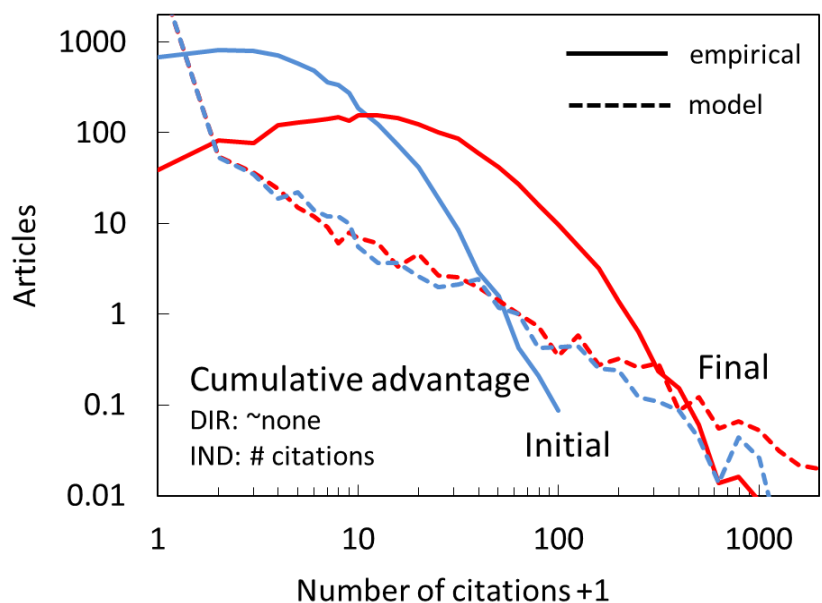

Figure 4. Citation distribution for the cumulative advantage model (dashed lines) compared to empirical distributions (full lines) after the first full year (Initial, blue) and at the end of the full period (Final, red), corresponding to years 2008 and 2017. In this model there is no direct citation mechanism (see text for details), whereas indirect citing is directly proportional to citations already accrued.

*3.4. Price model*

A conceptual and practical problem with the pure CA model described in the previous section is that in order for citations to accumulate by this mechanism there already need to exist some citations, which is not the case at the time when the paper is published. This problem of CA has already been recognized by Price [11], who refers to it as the *ground state* problem. In practical terms, we have addressed it by allowing some small fraction of direct citations to take place: $P \propto \epsilon + n_{cit}$, $\epsilon = 0.01$. This probability needs to be small so that most instances are still driven by CA. Obviously, this also means that most papers will never be cited, as can be seen in Figure 3.

In the context of generalized CA, in which the probability does not need to depend only on citations, the ground state problem can be "solved" by modifying the probability so that it is not zero when $n_{cit} = 0$. A popular way to achieve this modification, introduced by [11], is to assume the probability to be of the form:

$$P \propto 1 + n_{cit}. \qquad (3)$$

The implications of this simple modification are sometimes not fully appreciated, as it can be seen only as a practical way to get CA to operate. Instead, modifying the probability in this way is equivalent to introducing a *direct* citation mechanism that operates *independently* from CA. The deeper implication of such a modification is that a pure CA citation model is not possible, either practically or conceptually. Importantly form the standpoint of this work, adding one (or some other constant $\alpha$), to the probability introduces not just any direct mechanism, but one with a uniform probability. Price tentatively added one (rather than some other constant) as *de facto* considering the publication itself as its first citation. Our interpretation is more natural, since it ties $\alpha$ to a separate citation mechanism.

Price's model is known to be relatively successful in reproducing the empirical citation distributions. We show its predictions in Figure 5.

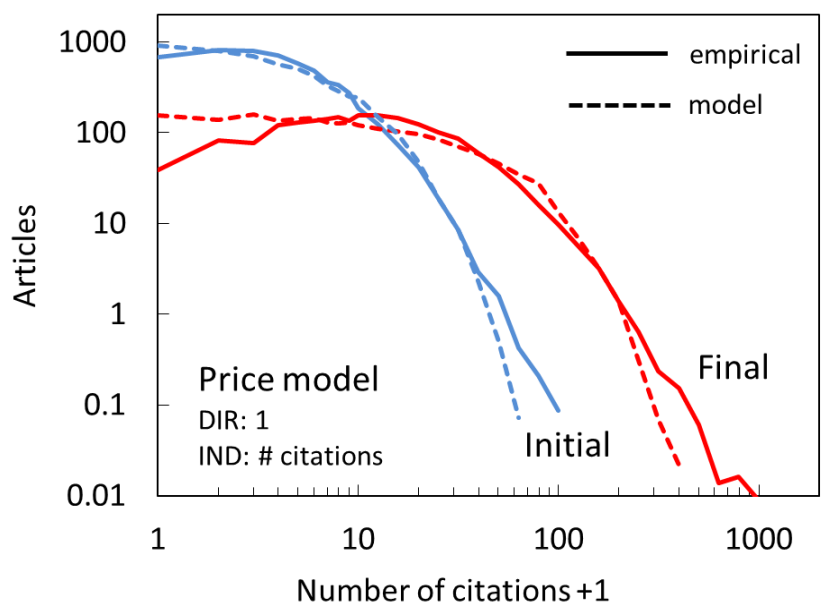

Figure 5. Citation distribution for the Price model (dashed lines), which implicitly combines direct citation with uniform probability and indirect citation based on cumulative advantage, compared to empirical distributions (full lines) after the first full year (Initial, blue) and at the end of the full period (Final, red), corresponding to years 2008 and 2017.

A general agreement is encouraging. Closer inspection reveals some important differences between model predictions and empirical distributions. The model peaks for papers with zero citations even for the final distribution, whereas the actual peaks have more citations. Furthermore, the drop in the tail is steeper in the model than in the actual distribution, i.e., not as many papers will have very large number of citations.

It needs to be pointed out that in our simulation the Price model is implemented so that the direct and indirect mechanisms operate concurrently. An alternative would be to only have direct mechanism operate at first, as long as needed to get the number of papers with zero citations to drop to the empirical number, and then allow the pure CA to take over. We have found such scenarios unrealistic. According to such a model, the phase of direct-only citations would dominate the initial distribution, and make it Poissonian. However, the empirical distribution already shows clear signatures of both mechanisms (i.e., a power-law tail).

*3.5. New, team-size based model*

The success of the Price model suggests that it may be possible to modify it to achieve an even better agreement with the empirical citation distributions. One such possibility is to modify the probability to be

$$P \propto \alpha + n_{cit}, \; \alpha \neq 1. \quad (4)$$

We refer to it as the generalized Price model. Price [11] already recognized that in principle $\alpha$ need not be 1, but has not provided a rationale for some alternative number, and instead considered it a "fudge" factor that may help get a better agreement with the empirical data. In the two-mechanism model $\alpha$ has a clear interpretation: it modifies the relative contributions of the direct and indirect mechanisms. In the limiting cases, if $\alpha \ll n_{cit}$ one gets pure CA, whereas if $\alpha \gg n_{cit}$ the model tends towards a uniform probability model (i.e., it becomes just the direct mechanism). Peterson et al. [20] have already noted what they refer to as the "similarity" between the generalized Price model and the two-mechanism model. We point out that the two are actually the same. The generalized Price model has its network theory equivalent in the modification of the preferential attachment to include a constant term, proposed by Dorogovtsev et al. [34] and referred by them as the initial attractiveness of a node.

We find that a simulation based on $\alpha \approx 1.5$ provides a better match with our empirical data - primarily by shifting the peak towards higher values compared to $= 1$. However, the shapes of model distributions do not match in detail, and the steep drop in the tail is still present. Is it possible to do better than that? One possibility, at least mathematically, is to allow the CA component to be sub ($\beta < 1$) or super ($\beta > 1$) linear [35, 36] resulting in the total probability of the form $P \propto \alpha + n_{cit}^{\beta}$. Such model has two free parameters. However, the justification for such modification from the standpoint of the processes by which researchers cite literature is unclear. To better match the empirical distributions in the tail, one needs to boost the CA by assuming superlinear dependence. It is not obvious why highly cited papers would have a visibility that exceeds the number of times they were cited. The sublinear scenario could be more justified (e.g., some fraction of citations may be perfunctory and does not transfer into a greater visibility of the cited paper), but such modification only increases discrepancies with respect to actual citation distributions. We confirm this by performing simulations and finding that neither sublinear nor superlinear dependence on the number of citations helps achieve a better agreement with the empirical distributions.

In this work, we propose to retain the CA mechanism in its simplest and most justified linear form (as already argued by Peterson et al. [20]), and instead modify the mechanism of direct citation. Rather than assuming direct citing with uniform probability, we propose to make it dependent on some other characteristic of the paper. This modification is rooted in the observation that there is no reason why all papers would have the same intrinsic citation capacity (by intrinsic we mean one that is not the result of the CA process), which then translates into initial visibility. Many factors can influence intrinsic citation capacity: prestige of the authors or the team, a hot topic, etc. These factors are difficult to quantify and incorporate in a simple citation model. However, one factor that is known to be a strong determinant of an overall citation performance is a *team size*. It is not unreasonable to assume that the prestige and the popularity of a topic can be, to some extent, reflected in the size of the team.

From the modeling standpoint, the article team size is easy to measure by counting the number of authors on a paper. The knowledge production in astronomy, like in many other experimental fields takes place in teams of varying sizes. Specifically, for our dataset consisting of 6430 articles published in 2007, the team size distribution is shown on a log-log plot in Figure 6. Teams range from a single author to several hundred authors, without any gaps in between. The majority of papers are still produced by smaller teams. Most common team size consists of 2 and 3 authors. Team size distribution features a peak and a power-law like tail, indicating multiple formation mechanisms [37].

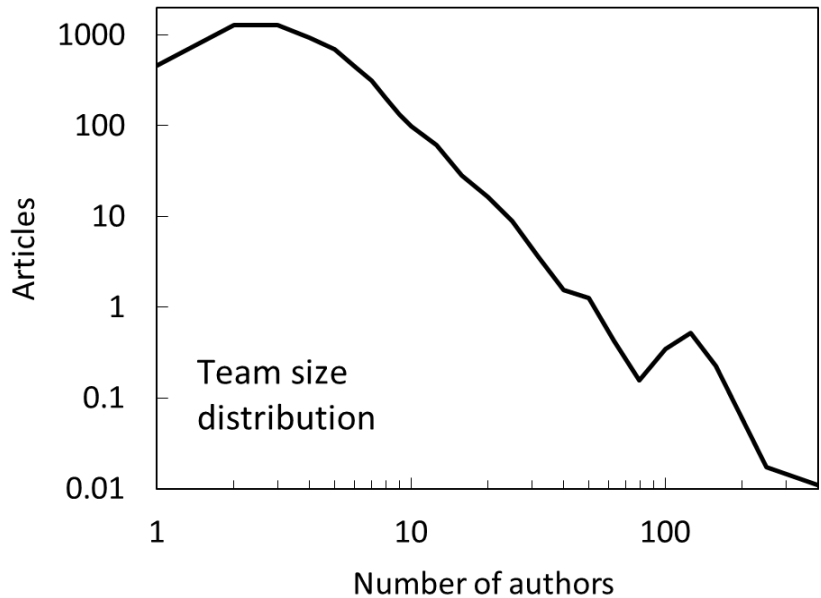

Figure 6. Team-size distribution for astronomy articles published in 2007. We propose that the probability of direct citation is proportional to the number of authors, except for very large teams, where it stays constant.

Based on the above considerations we test the performance of the model in which the probability of direct citation is proportional to the paper team size measured by the number of authors, whereas the indirect citation follows CA in its simplest linear form:

$$P \propto n_{author} + n_{cit}. \qquad (5)$$

The results of the application of this two-mechanism model are shown in Figure 7. The agreement for both the initial and the final citation distribution is excellent, with a standard deviation between model and empirical binned values of only ~0.1 decades. It needs to be pointed out that in applying the model we treated all papers with more than 30 authors as if they had exactly 30 authors. This capping is based on the fact that the citation benefit does not increase past some team size (30 for this dataset; Figure 8). The reason for that may be that very large teams list as authors the team members that contribute to various aspects of the project but are not directly involved in the knowledge production (i.e., the support scientists [38]) and therefore do not drive citation dynamics.

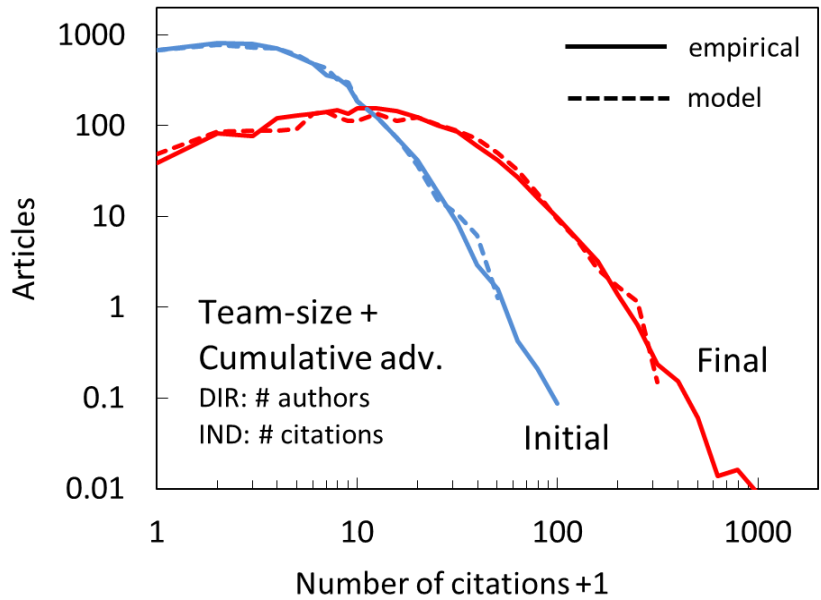

Figure 7. Citation distribution for the new, two-mechanism model (dashed lines), in which direct citation has a probability proportional to the paper team size (number of authors) and indirect citation is based on cumulative advantage, compared to empirical distributions (full lines) after the first full year (Initial, blue) and at the end of the full period (Final, red), corresponding to years 2008 and 2017.

The simulation based on our model allows us to determine the relative contribution of direct and indirect citations over time. The results are given in Figure 8 as the percentage of citations that can be attributed to direct citation among the citations received in some year. The model predicts a steady decline in the contribution of direct citations. From 80% in the year in which the papers were published, to ~10% in later years. This drop is expected since, as the time passes, the chances of learning about the paper from some other paper rather than directly will increase.

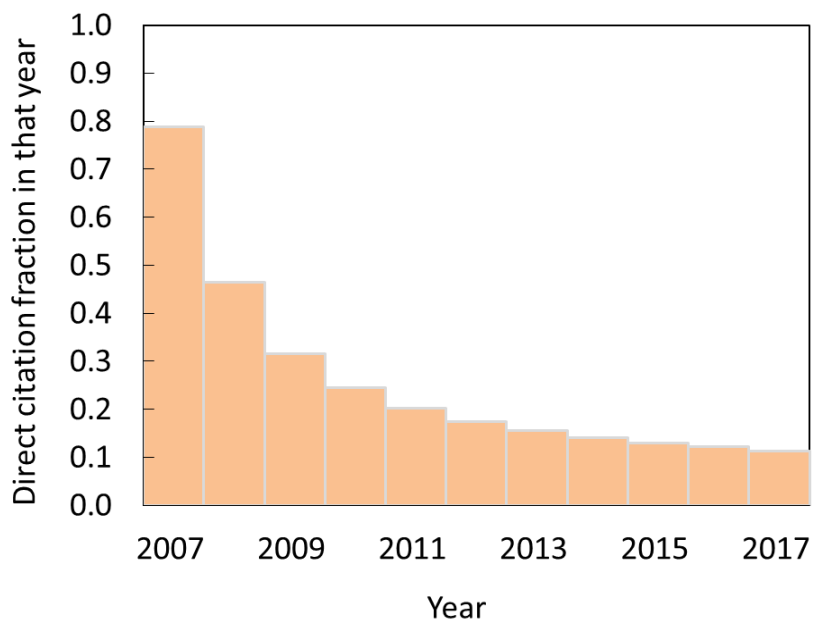

Figure 8. The change of the share of direct citations among the citations received by astronomy articles published in 2007 in subsequent years/ The numbers are obtained by simulating the citation processes according to our preferred two-mechanism model (Fig. 7).

Also based on the model we can explore the contribution of the two mechanisms for articles that have a certain number of citations at the end of the period (Figure 9). As expected, the fraction of citations obtained directly drops for papers with greater number of citations. The break-even point is for papers with ~7 citations. Interestingly, the fraction remains significant even for most-cited papers and seems to plateau around 25%.

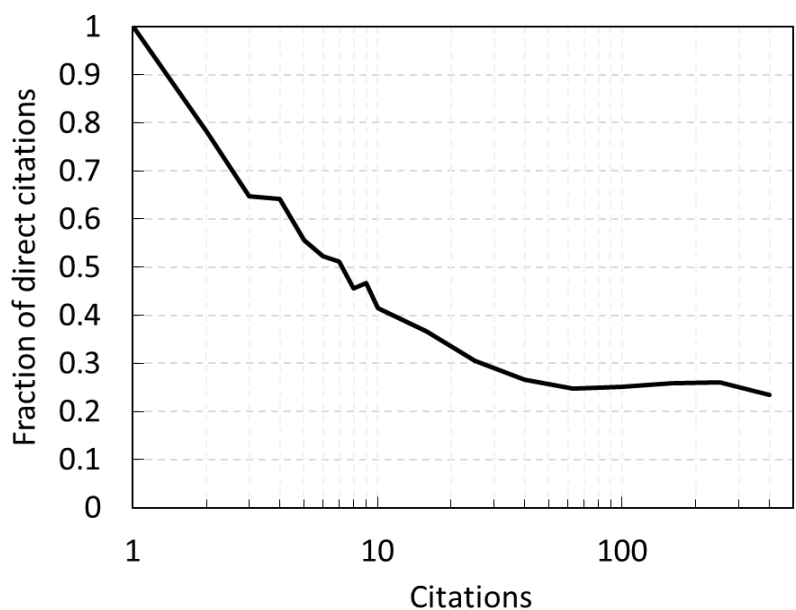

Figure 9. Fraction of direct citations for papers with different number of total citations at the end of the citation period (end of 2017.

Finally, we point out that our model can be generalized as:

$$P \propto f(n_{author}) + n_{cit}. \qquad (6)$$

However, at least for our dataset, we have found that $f(n_{author}) = n_{author}$ gives better results than simple linear or power-law modifications.

## 4. Discussion

*4.1. The significance of our two-mechanism citation model with non-uniform direct citation*

Simulations presented in the previous section demonstrate that all successful citation models have to assume two citation mechanisms, whether this is explicitly recognized, or, as in the case of the Price model, the consequence of a practical solution to the initialization (ground-state) problem required for cumulative advantage process. The reasons why the probability for indirect citing would scale with the number of citations and therefore lead to a cumulative advantage are fairly straightforward and we have already discussed them. The reasons why the probability of direct citation would scale with the team size in our model may be less obvious. As we discussed, direct citation is based on the idea that a researcher has learned about a paper in ways other than seeing it cited in some other work. Researchers who will for certain be aware of a paper in a direct way are its authors. Consequently, the larger the team size the greater the chances that some of its authors will subsequently cite that work. It is therefore not surprising that direct citation would be related to team size for this reason alone, in addition to, as we remarked, often greater initial visibility of works by greater teams due to their previous contributions.

That the probability of direct citation is not uniform has been anticipated already by Price [11]. Even though he did not realize the existence of two citations mechanisms and has not considered $\alpha$ that wouldn't be constant, Price [11] speculated about there being a *range* of "initial citation pulses". He suggested that this pulse may be related to the "quality" (his quotes) of the paper, and could perhaps include some easily quantifiable characteristic of a paper such as its length in pages. It is important to point out that in the two-mechanism interpretation, this "initial pulse" is actually present at all times in the form of direct citation, which is why we prefer to call it the "intrinsic" citation potential, rather than the "initial". Naturally, over time, as we will see next, the citations accrued indirectly will outnumber the direct citations, but for papers with few citations this "initial" phase will take a very long time. Furthermore, we show that in order to obtain the empirical characteristics, this direct citation probability needs to be proportional to some quantity that itself has a wide range (such as the team size distribution), rather than the length of the paper, which is distributed more or less normally.

Golosovsky & Solomon [28] have shown that the distribution of the accumulation rate of citations is not random because of the temporal correlations, such that a paper that is "on a roll" will likely continue to be cited at a higher rate. They claim that the inclusion of this effect into a preferential attachment model, which they call multiplicative stochastic model, leads to citation distributions that more accurately follow the empirical distribution than just the preferential attachment model in which the probability of direct citation (what they call "initial attractivity") is uniform. However, we find that similar improvements are possible by making direct citation not have uniform probability. In fact, the nature of the modifications in two models is similar, except that we use an external parameter (number of authors), whereas Golosovsky & Solomon need to consult empirical citation data to determine the initial citation rate, i.e., they effectively rely on extrapolation. Consequently, our interpretation is not that the cumulative advantage process itself is too rigid, but rather that it needs to operate on article-specific, heterogeneous "initial attractivity" (what we call intrinsic citation capacity) in order to achieve the required flexibility.

*4.2. Do we really need both mechanisms?*

The team size distribution itself has a power-law tail, as can be seen in Figure 6, which [37] explain to be due to the principles by which some teams grow that are themselves rooted in the process of cumulative advantage. One may wonder whether such team size distribution is alone the reason why our model successfully predicts the empirical distribution, without a need for a CA. We test this by performing a simulation in which $P \propto n_{author}$, i.e., that lacks an indirect citation mechanism. The results are shown in Figure 10. We see that the initial distribution is reproduced reasonably well, however, the final distribution is much narrower and severely underpredicts the number of articles with <10 citations. In other words, we confirm that CA is essential in producing broad citation distributions with large disparities in the number of citations.

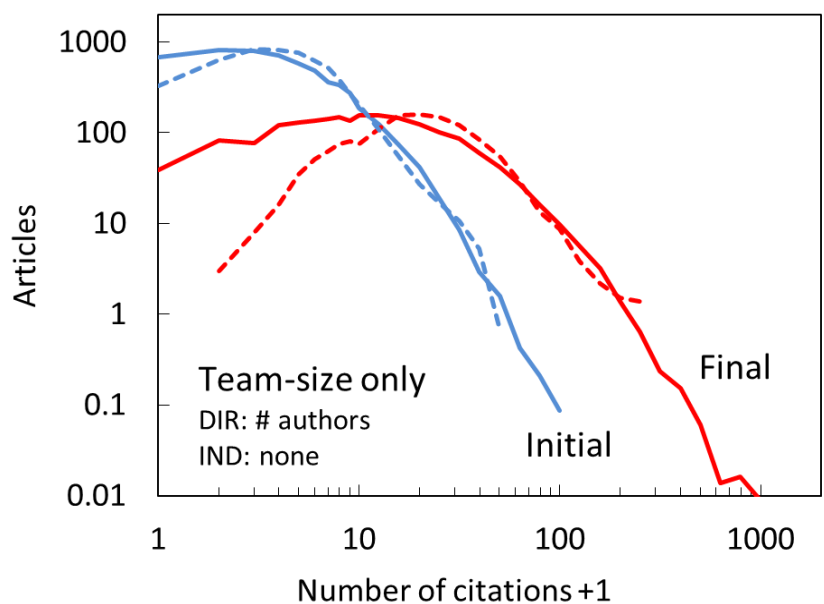

Figure 10. Citation distribution for model (dashed lines) which only has direct citation mechanism, based on the team size. The agreement with empirical distributions (full lines) is worse than in our preferred, two-mechanism model that also includes cumulative advantage (Fig. 7), especially for the final distribution.

*4.3. The scope and limitations of the current model*

In this paper we focus on citation distribution as the principal benchmark for the success of a model. One can in principle test if and how well the model reproduces other empirical properties. As mentioned in the introduction, there is a well-known correlation between paper team size and the number of citations received. In our model, we explicitly require the probability of direct citation to scale with team size. To see if this connection is preserved in overall citations where indirect citations make only a smaller fraction of citations, in Figure 11 we show the average number of citations as a function of team size for the model (dashed line) and actual citations (solid line). We see that the model gives a stronger than actual trend. The correlation between the number of citations and the team size can be very well reproduced if we generalize our model as $P \propto cn_{author}^{\gamma} + n_{cit}$, (a specific form of Eq. 6) and choose $\gamma = 0.3$ and $c = 1$, i.e., assume a weaker, sublinear dependence on the number of authors. This results in a relation shown in Figure 11 as a dot-dashed line. Would this modification be a better alternative to our preferred model? Unfortunately, while the modified model agrees with the empirical citation distributions relatively well (better than the Price model) the preferred model is clearly superior.

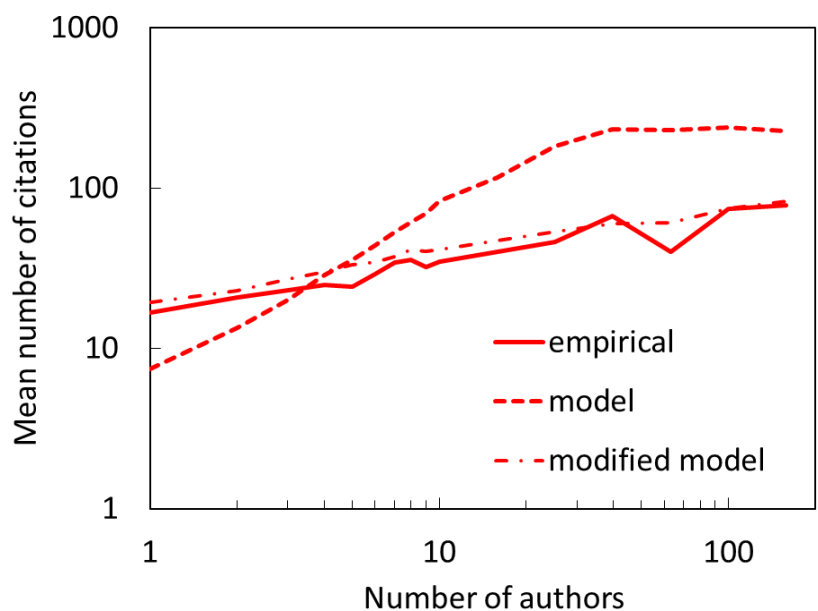

Figure 11. Mean number of citations received by papers with different number of authors (team size). Our preferred two-mechanism model (dashed line) gives a stronger relation than what is found in the empirical data (full line). The model can be modified by making the direct citation more weakly dependent on the number of authors (dot-dashed line), however, this comes at the expense of somewhat inferior ability of the modified model to reproduce the empirical citation distribution. Citations are for the end of the period (2018). The mean is a geometric mean.

Our team sized model has been tested and makes sense for regular research articles, not for items such as review articles. Their intrinsic citation capacity cannot be related to the usually small number of authors of review papers, and is instead more likely to depend on other factors.

As mentioned in previous section, we find that the linear dependence of the probability of direct citation on team size reproduces the empirical data better than a generalized form with one or two extra parameters. Nevertheless, the team-size dependence may very well be field dependent with the exponent $\gamma$ and multiplicative constant $c$ taking values other than one. It is therefore recommended to test and eventually adjust our model for different fields or research areas separately.

Finally, we comment on the temporal aspect of citation dynamics, which we consider separate from the citation mechanisms. Wu and Holme [27] and Eon and Fortunato [18] developed citation models using network approach and defined a kernel function consisting of two terms, which correspond to our direct and indirect citation mechanisms, even though they do not describe their models in those terms, and [18], following [34], refer to the first term as *attractiveness*. Both models allow the probability of direct citation to decline in time. In our model, direct citation has no dependence on time and yet we can reproduce the citation distributions of a cohort of papers at various points in time very well. The difference arises from the fact that unlike us they consider the citation of a body of literature published over long time intervals, so their papers have, at any given time, a range of ages. The obsolescence of scientific literature is widely known and is a separate question from the citation mechanism. We argue that it is conceptually easier to study the citation dynamics (i.e., the obsolescence) separately from the citation mechanisms. In our approach, all articles have the same age that increases as time goes by and new citations are distributed among them according to a citation model as long as those papers are cited, regardless of whether the overall citation declines over time, stays the same, experiences bursts, etc.

*4.4. Future directions*

All cumulative advantage models including our team-size model use as their basis just the *number* of existing citations, i.e., they assume that each previous citation provides the same advantage for subsequent citation. In reality, it would be more reasonable to expect that some citations will confer greater cumulative advantage effect than the others. Being cited by an influential paper intuitively brings greater visibility than being cited by an obscure one. In such case our model could be modified such that the citation probability becomes

$$P \propto f(n_{author}) + \frac{1}{\langle g(I) \rangle} \sum_{i=1}^{n_{cit}} g(I_i), \tag{7}$$

where $I$ is the influence of each citing paper, for example, the number of citations in the first few years after its publication. Our initial tests of this model fail to demonstrate clear improvements, to some extent because the simpler model already fits the empirical citation distributions so well.

In closing, we point out that we do not claim that the size of the team is the only or even the principal determinant of its citation potential. Rather, we argue that to produce the correct citation distribution for an *ensemble* of articles that has appeared in journals of similar overall prestige, the citation potential needs to be distributed in a similar way as the team size distribution (having a power-law tail). Interestingly, this agrees with the assumption of [18] that the intrinsic citation potential (what they call initial attractiveness, $A_0$) is drawn from some arbitrary power-law distribution with an exponent $-2.5$. Intrinsic citation potential is not a quantity that can be easily measured or even well defined, but the existence of the correlation between the total citation received and the team size suggests that the team size may serve as a good proxy, and our simulations confirm that such assumption may be sufficient. The citation potential should be expected to depend on the initial visibility, and having more people be intimately involved and familiar with a paper provides that in a natural way. Even if we had a way to define and measure a "true quality" of a paper this may not be the best quantity to use in the model, as the citation trajectory of a paper will still depend on its visibility, which larger teams are more likely to provide. Clearly, our team-size based model cannot be applied to datasets where the majority of papers are authored by single authors or very small teams, as would correspond to the literature from older periods or some fields even today. In those cases it may be harder to find an alternative proxy for initial visibility that would not be in one form or another based on the knowledge of the previous success of the authors.

## 5. Conclusions

Although more work is needed to arrive at the model that would be able to predict all empirical distributions and relations equally well, this work has set a solid foundation for that effort by demonstrating that it is essential to include both direct and indirect citation mechanisms, and that the former may be related to a property (or properties) of the paper that reflect its intrinsic citation potential, and therefore does not have a uniform probability. Our conclusions are listed below.

- Our simulations confirm that the modeling of empirical citation distributions requires more than one citation mechanism, as suggested by [25].
- The two mechanisms correspond to direct and indirect citation, as proposed by [20, 26]. Furthermore, we interpret direct mechanism as the intrinsic citation capacity of a paper that governs its initial visibility. Indirect mechanism consists in learning about a paper when it is cited in another paper and is subject to cumulative advantage.
- Our simulations show that the role of direct citation is not only to provide initial citations for the indirect mechanism of cumulative advantage to be able to operate. Rather, it operates concurrently with indirect citation.
- Over time, the majority of new citations accrue via the indirect method, as pointed out in e.g. [18].
- The addition of 1 in Price's citation model [11], originally motivated by the need to resolve the ground-state problem (that citations are initially zero) is actually equivalent to introducing a

- direct citation mechanism with uniform probability, which solves ground-state problem in a natural way.
- The meaning of the constant in generalized Price model [11] (constant other than one) is that it determines the relative strength of direct vs. indirect citation.
- Even the generalized Price model [11] does not reproduce the tail of citation distribution well, underpredicting the number of papers with very high citations.
- Critically, we show that a two-mechanism model in which direct citation probability is not uniform, but instead draws from a broad distribution, is needed to reproduce the empirical distribution in detail, including its tail as well as the number of papers with no citations.
- We demonstrate that a two-mechanism model in which the probability of direct citation is proportional to the number of authors on a paper (team size) reproduces the empirical distributions remarkably well. This model is proposed and tested for regular articles, not review papers where the intrinsic citation capacity is unlikely to be related to usually small author list. We do not claim that the intrinsic citation capacity depends only on the team size, but is to some degree correlated with it and is distributed in a similar way, having a power-law tail.
- Direct citation, even when based on team sizes, cannot produce extreme disparities in citation counts alone.
- Our team-size model qualitatively explains the existence of a correlation between the number of citations and the number of authors on a paper.
- Interpretation of our model is that intrinsic citation capacity will be greater the more people are intimately familiar with some work, favoring papers from larger teams, up to ~30 authors, for our dataset.

While the principles behind the proposed model are expected to be universal, future work will test the model in other disciplines and over different time periods.

**Funding:** This material is based upon work supported by the Air Force Office of Scientific Research under award number FA9550-19-1-0391.

**Acknowledgments:** This work uses Web of Science data by Clarivate Analytics provided by the Indiana University Network Science Institute and the Cyberinfrastructure for Network Science Center at Indiana University.

**Conflicts of Interest:** The author declares no conflict of interest.